\documentclass[conference]{IEEEtran}
\IEEEoverridecommandlockouts
\usepackage{amsmath,amssymb,amsfonts}
\usepackage{algorithmic}
\usepackage{graphicx}
\usepackage{textcomp}
\usepackage{xcolor}

\usepackage{url}

\usepackage[hidelinks,breaklinks]{hyperref}
\usepackage{breakurl}

\def\BibTeX{{\rm B\kern-.05em{\sc i\kern-.025em b}\kern-.08em
    T\kern-.1667em\lower.7ex\hbox{E}\kern-.125emX}}

\usepackage{xtab,booktabs} 

\usepackage{graphicx}
\usepackage{subfig}

\makeatletter
\newcommand\footnoteref[1]{\protected@xdef\@thefnmark{\ref{#1}}\@footnotemark}
\makeatother

\usepackage{biblatex} 
\begin{document}

\title{An Enhanced Approach to Cloud-based Privacy-preserving Benchmarking (Long Version)}

\author{\IEEEauthorblockN{Kilian Becher}
	\IEEEauthorblockA{\textit{Chair of Privacy and Data Security}\\
		\textit{TU Dresden}\\
		Dresden, Germany\\
		kilian.becher@tu-dresden.de}
	\and
	\IEEEauthorblockN{Martin Beck}
	\IEEEauthorblockA{\textit{Chair of Privacy and Data Security}\\
		\textit{TU Dresden}\\
		Dresden, Germany\\
		martin.beck1@tu-dresden.de}
	\and
	\IEEEauthorblockN{Thorsten Strufe}
	\IEEEauthorblockA{\textit{Chair of Privacy and Data Security}\\
		\textit{TU Dresden}\\
		Dresden, Germany\\
		thorsten.strufe@tu-dresden.de}}

\maketitle
\thispagestyle{plain}
\pagestyle{plain}

\begin{abstract}
	Benchmarking is an important measure for companies to investigate their performance and to increase efficiency. As companies usually are reluctant to provide their key performance indicators (KPIs) for public benchmarks, privacy-preserving benchmarking systems are required.
	In this paper, we present an enhanced privacy-preserving benchmarking protocol, which we implemented and evaluated based on the real-world scenario of product cost optimisation. It is based on homomorphic encryption and enables cloud-based KPI comparison, providing a variety of statistical measures.
	The theoretical and empirical evaluation of our benchmarking system underlines its practicability.
\end{abstract}


\section{Introduction}\label{Introduction}

Benchmarking is the comparison of key performance indicators (KPIs) of a company's peer group~\cite{KER08}. KPIs are statistical quantities that can be used for evaluating the performance of a company~\cite{KER08}.
A peer group is a set of companies that take advantage of comparing KPIs~\cite{KER08}. Usually, the companies of a peer group are competitors of the same industry, which implies a demand for privacy of the KPIs~\cite{KER08}.
We define a privacy-preserving benchmarking analysis as the process of comparing KPIs as secure inputs across different companies~\cite{CDN15}. Every company learns how it performs compared to the other companies being involved but not their private KPIs~\cite{CDN15}.

One approach to privacy-preserving benchmarking is using a trusted third party (TTP) that conducts the calculation of a function $f(x)$ without revealing any private data. However, with mutually distrusting companies, finding a TTP might not be trivial~\cite{GMW87}. An approach that does not require trust can be found in secure multi-party computation (MPC).
In MPC, participants providing an input $x_i$ are called players, those who compute $f(x)$ are called processors~\cite{HIR01}. A participant can be both player and processor at the same time.
An MPC is secure in the sense that the participants only learn the outputs, their own input, and what they can infer from that~\cite{CDN15}.

In this paper, we investigate approaches to adding privacy-\linebreak preserving benchmarking to a product costing suite of the software company SAP.
Such product costing software enables a company to calculate the costs of its products.
This includes quotation costing as well as cost estimation from the project acquisition to the design and production to the disposal. 

To increase profit, a company, e.g. of the automotive industry, might want to reduce production costs. To do so, it needs to determine those areas with the best ratio between cost savings and optimisation effort, requiring knowledge of the company's performance. To make well-informed optimisation decisions, optimisation can be facilitated by comparing the company's production KPIs, e.g. assembly time of a car engine, to those of other companies of the industry, i.e. via benchmarking analyses. If according to the benchmarking results a company finds itself among the best performing, it might be rather expensive to further improve the compared KPI, e.g. assembly time. A performance below average might imply higher potential for cost savings. However, as companies might be reluctant to provide their confidential production KPIs, benchmarking needs to be conducted in a manner that ensures privacy of the KPIs and still provides the desired statistical measures. Such privacy-preserving benchmarking analyses could be repeated on a regular basis, e.g. once every quarter, to investigate performance development over time relatively to the industry.

The requirements for our system were defined given the corporate context. They are based on the requirements of an existing, TTP-based benchmarking suite of SAP, serving as a lower bound for functionality and performance. 
In this TTP-based system, benchmarks are conducted individually for each KPI at least once every six to seven months on pre-defined dates rather than on demand. It provides the statistical measures mean, bottom quartile, top quartile, and best-in-class.
The latter is the mean of the top quarter of the sorted list of inputs. As in the existing benchmarking system and similar to Atallah et al. in~\cite{ABL04}, we assume companies comparing their product costing KPIs to be interested in correct results and therefore to behave honestly. This can reasonably be assumed as companies need such benchmarking results for well-informed, far-reaching business decisions. These results are only correct if every participant follows the protocol. Hence, we require our system to provide input privacy against semi-honest adversaries. To guarantee that each company gets the same, correct results, output integrity is required. Similar to the TTP-based system, our system should provide anonymity among the players in the sense that participants must not be referred to with any persistent identifier during benchmarking~\cite{KER10}. This might be of interest in case of computing benchmarks only for a subset of a peer group. Further aspects like availability were beyond the scope of our considerations.

Following the company's cloud strategy, our system should run as a cloud service.
This provides scalability and availability and can be assumed to enable protocol execution even for large peer groups~\cite{AFG10}. In the TTP-based system, such large peer groups contain up to $300$ players, which is sufficient for the product costing context.
As suggested in Kerschbaum's guideline for non-functional requirements of a benchmarking platform~\cite{KER10}, a benchmark should at most take $24$ hours.

Our main contributions are the construction and implementation of a privacy-preserving benchmarking system in the context of product cost optimisation. Furthermore, we conducted security and complexity analyses and an extensive performance evaluation of this protocol in the given context.

Focusing on the technical feasibility of privacy-preserving benchmarking systems, we first performed a state-of-the-art analysis for possible approaches (Section~\ref{Approaches}). We then selected one approach, adapted it to better suit the product costing context, designed a prototype (Section~\ref{Design}), and evaluated it extensively regarding the given requirements (Section~\ref{Evaluation}).

\section{Preliminaries}\label{Preliminaries}

\subsection{Oblivious transfer}\label{PreliminariesPrimitivesOT}

In an oblivious transfer (OT) protocol, a player $P_1$ has $k$ secret messages $m_1,...,m_k$ with $k \geq 2$. A player $P_2$ wants to select and receive message $m_i$ without $P_1$ learning the value $i$~\cite{KER10}. Furthermore, $P_1$ does not want $P_2$ to learn any message apart from $m_i$~\cite{KER10}. We denote such a protocol by $P_1 \xrightarrow{\mathrm{OT}} P_2$.

\subsection{Homomorphic encryption}\label{PreliminariesHomomorphicEncryption}

Assume a cryptosystem with a (randomised) encryption function $E(\cdot)$ and a decryption function $D(\cdot)$~\cite{KAL08}. Homomorphic cryptosystems enable computations for secret values $x_1,...,x_n$ based on their ciphertexts $E(x_i),...,E(x_n)$ without needing the decryption key~\cite{KER08}. Applying an operation to such ciphertexts yields the ciphertext of the result of a corresponding homomorphic operation applied to the plaintexts~\cite{KER08}. 
Partially homomorphic encryption (PHE) schemes enable one operation on the plaintext, e.g. addition or multiplication~\cite{KER10}. Fully homomorphic encryption (FHE) schemes allow for the computation of arbitrary functions, e.g. by providing both addition and multiplication~\cite{GEN09b}. For example, Paillier's asymmetric, additively (partially) homomorphic encryption scheme has the properties given in Equations~(\ref{eq:PreliminariesHomomorphicEncryptionAdditionMult}) and~(\ref{eq:PreliminariesHomomorphicEncryptionAdditionExp})~\cite{KER08}.
\begin{equation}\label{eq:PreliminariesHomomorphicEncryptionAdditionMult}
D\left(E\left(x_1\right) \cdot E\left(x_2\right)\right) = x_1 + x_2
\end{equation}
\begin{equation}\label{eq:PreliminariesHomomorphicEncryptionAdditionExp}
D\left(E\left(x_1\right)^{x_2}\right) = x_1 \cdot x_2
\end{equation}
Homomorphic semantically secure cryptosystems provide rerandomisation of ciphertexts as follows~\cite{KER10}.
\begin{equation}\label{eq:PreliminariesHomomorphicEncryptionRerandomisationAddition}
E(x_i + 0) = E(x_i) \cdot E(0) = E^\prime(x_i)
\end{equation}
With high probability, $E(x_i) \neq E^\prime(x_i)$ is provided such that $E(x_i)$ and $E^\prime(x_i)$ are computationally indistinguishable~\cite{KER10}.

\section{Related Work}\label{Approaches}

We assess the suitability of several existing approaches and protocols for privacy-preserving benchmarking regarding the requirements described above. It covers the generic approaches of garbled circuits, secret sharing, and homomorphic encryption as well as custom protocols. The former can be used for computing arbitrary functions while the custom protocols only enable computation of predefined functions.

\subsection{Garbled circuits}\label{ApproachesBMR90}

The protocol presented by Beaver, Micali, and Rogaway in~\cite{BMR90} implements the garbled circuits approach for multiple players.
The players create a Boolean circuit that implements $f(x)$.
Even though one might consider garbled circuits to be impractical~\cite{HEK11}, it was shown that they can compete with custom protocols regarding efficiency~\cite{HEK12}. These findings were driven by improvements of the garbled circuits approach, such as free XOR and garbled row reduction (see~\cite{HEK11,KAS08,PSS09}). For specific tasks, e.g. calculation of the mean, one only needs to create a garbled circuit that implements the corresponding operations instead of developing an entire protocol.

The protocol of~\cite{BMR90} consists of two phases, one for jointly generating a common garbled circuit $C$ together with the garbled input and a second for publishing and evaluating this circuit. Since the protocol implements the generic approach of garbled circuits, it can be used for calculating any computable function~\cite{BMR90}. However, up-front effort is required to create the necessary circuits.
The most important drawback of this approach is its non-centralised communication model. It requires pairwise communication between the players, which precludes inherent anonymity among them. 
Therefore, the approach presented in~\cite{BMR90} does not suit the requirements properly.

\subsection{Linear secret sharing and homomorphic encryption}\label{ApproachesLinearSecretSharingHomomorphicEncryption}

This approach is a combination of linear secret sharing and homomorphic encryption. The idea of secret sharing is to split the secret values into shares and spread these shares among the players involved in the scheme~\cite{CDN15}. The function $f(x)$ for these inputs can then be computed given the shares~\cite{CDN15}.
In this approach, the players $P_i$ share their secret input values $x_i$ among the set of $n$ players using a linear secret sharing scheme like Shamir's scheme (see~\cite{SHA79}). Each player $P_j$ holds one share $[\![ x_i ]\!]_j$ for each of the $n$ secret-shared values $[\![ x_i ]\!]$. Addition of shares is a linear operation and can be performed locally while multiplication requires a subprotocol that introduces overhead~\cite{HIR01}. Such a subprotocol for two players is presented by Atallah et al. in~\cite{ABL04}. The complexity of this subprotocol is exponential in $n$.

In its original form, the approach has a non-centralised communication model requiring pairwise communication between the players. Therefore, the anonymity requirement is not met.
The most important drawback of this approach is the complexity of the multiplication subprotocol making it impractical for large $n$, i.e. large peer groups.
Consequently, the approach described in this Section does not suit the requirements properly.

\subsection{Fully homomorphic encryption}\label{ApproachesFullyHomomorphicEncryption}

Given a FHE scheme with an encryption function $E(\cdot)$ and the encrypted secret values $E(x_1),...,E(x_n)$, one can compute the encrypted result $E(y)=E(f(x_1,...,x_n))$ for any efficiently computable function $f(x)$ without needing to decrypt~\cite{GEN09b}.
Given that, one can build a secure MPC system where one player conducts the entire computation of $E(f(x))$ locally without learning anything about $x_1,...,x_n$ or $y$.

In~\cite{GEN09b}, Gentry presents the first ever FHE scheme. This seminal work started a new research field and led to many significant improvements. 
However, even enhanced approaches to FHE schemes, such as the one presented by Brakerski, Gentry, and Vaikuntanathan in~\cite{BGV14}, have complexity at least polynomial in the size of the respective circuit with large constants. Therefore, this approach can reasonably be assumed to not suit the performance requirements properly.

\subsection[Custom protocol]{Custom benchmarking protocol}\label{ApproachesKER10}

In~\cite{KER10}, Kerschbaum presents a secure MPC protocol designed for privacy-preserving benchmarking platforms that has constant cost\footnote{Constant cost here means constant round complexity and constant, i.e. linear in the size of the security parameter $\kappa$, communication complexity. Both are independent of the peer group size~\cite{KER08}.}, provides anonymity among the players, and is centralised.
It requires a single server; the service provider $P_S$ of the benchmarking platform who acts as a processor.
The protocol enables the benchmarking platform to compute the statistical measures mean, variance, median, maximum, and best-in-class~\cite{KER10}. The protocol is based on an additively partially homomorphic encryption scheme.

The protocol natively implements the service provider model and therefore can ensure anonymity among the players as well as cloud suitability.
Input privacy is ensured in the semi-honest model as well as in the constrained malicious model~\cite{KER10}. Output integrity is ensured via message authentication codes (MACs).
With a fixed set of four rounds, its round complexity is constant.
However, the protocol only offers a subset of the required statistical measures.
The protocol's computational and communication complexity both are quadratic in $n$. This may prove critical for large peer groups.

\subsection[Custom aggregation protocol]{Custom aggregation protocol}\label{ApproachesBIK16}

In~\cite{BIK16}, Bonawitz et al. present a protocol for secure aggregation.
This protocol assumes two kinds of participants: one server $P_S$, acting as a processor, and a set of $n$ players $P_i$, each providing a secret input $x_i$.
Players only communicate with the server $P_S$, who acts as a mediator between players.
Only $P_S$ learns the output $y=\sum_{i=1}^n x_i$, i.e. the sum of the secret values. The players do not learn anything new~\cite{BIK16}.

Executed in a central server scenario, the protocol enables\linebreak anonymity among the players as well as cloud suitability.
Rank-based statistical measures such as quartiles cannot directly be computed using this protocol. Even though its round complexity is constant, the protocol has computational and communication complexity that is quadratic in the number of players $n$, which may prove critical for large $n$~\cite{BIK16}.

\subsection{Summary and selection of a suitable approach}\label{ApproachesSummary}

The three generic approaches garbled circuits, secret sharing, and (fully) homomorphic encryption do not meet the requirements mostly due to their complexity. On top of that, their implementation in a centralised communication model would require additional effort and further increase the complexity. Otherwise, they would not meet the required level of anonymity.
The less complex protocol presented in~\cite{BIK16} is not suitable due to the lack of rank-based statistical measures. 
The most suitable approach is the privacy-preserving benchmarking protocol presented in~\cite{KER10}. 
Even though its computational and communication complexity of $\mathcal{O}(n^2)$ may prove critical for large peer groups, we build upon this protocol.

\section{Design}\label{Design}

The privacy-preserving benchmarking protocol of~\cite{KER10} (see Section~\ref{ApproachesKER10}) only offers a subset of the required statistical measures, i.e. mean, variance, median, maximum, and best-in-class. To provide the additional statistical measures bottom quartile $bq$ and top quartile $tq$, the protocol had to be enhanced. The full adapted protocol is given in this Section.

Prior to the protocol execution, each player $P_i$ learns the following two keys, e.g. with the help of a certificate authority as described in~\cite{KER10}.
\begin{itemize}
	\item $K_{DEC}$: The secret decryption key of the PHE scheme.
	\item $K_{MAC}$: The symmetric key of the MAC.
\end{itemize}
Every participant, including the service provider $P_S$, also learns the public encryption key $K_{ENC}$ corresponding to $K_{DEC}$. The players use the same secret key for decryption. They directly communicate only with the service provider, via pairwise channels that are secured based on standard methods for protecting transmission over insecure networks~\cite{KER10}.

\subsection{Adapted protocol}\label{DesignAdaptedProtocol}

Both the original protocol and our enhanced version are combinations of the techniques summation, rank computation, selection, and decryption~\cite{KER10}, which will be introduced first.

Summation of encrypted values is conducted by multiplying the ciphertexts (see Equation~(\ref{eq:PreliminariesHomomorphicEncryptionAdditionMult}))~\cite{KER10}. For $n$ values $x_i$, the encrypted sum is
\begin{equation}\label{eq:ApproachesKER10Summation}
E\left(sum\right) = E\left(\sum^n_{i=1} x_i\right) = \prod^n_{i=1}E\left(x_i\right).
\end{equation}
Summation is required for calculation of the mean $mean$ (steps 1 and 2) and of the variance $var$ (steps 13 and 14). The sum is blinded by adding a random value~\cite{KER10}. Since the players know the size $n$ of the peer group, each player can compute the mean himself by dividing the sum by $n$~\cite{KER10}.

Rank computation yields the rank of a value $x_i$ in a list which is sorted in ascending order~\cite{KER10}. To achieve that, the value $x_i$ is compared to each value $x_j$. For that comparison, the indices of the secret values are permuted by the permutations $\phi$ and $\phi^\prime$~\cite{KER10}. The assigned element of $i$ is denoted by $\phi(i)$ while the one of $j$ has index $\phi^\prime(j)$. The difference between $x_{\phi(i)}$ and $x_{\phi^\prime(j)}$ is blinded by two random values $1 \leq r_{2_j}$ and $0 \leq r_{3_j} \ll r_{2_j}$~\cite{KER10}. These are chosen individually for each $j$. The blinded difference
\begin{equation}\label{eq:ApproachesKER10RankDifference}
c_{\phi(i)_{\phi^\prime(j)}} = r_{2_j} \cdot \left(x_{\phi(i)} - x_{\phi^\prime(j)}\right) + r_{3_j}
\end{equation}
is stored in the vector $\vec {c_{\phi(i)}}$. Counting the non-negative elements $pos(\vec {c_{\phi(i)}})$ of that vector yields the number of input values that are smaller than $x_{\phi(i)}$~\cite{KER10}. Given that, its rank is
\begin{equation}\label{eq:ApproachesKER10Rank}
rank_{\phi\left(i\right)} = pos\left(\vec {c_{\phi\left(i\right)}}\right) + 1.
\end{equation}
Now, due to the permutations, each player $P_i$ holds the rank of the value $x_{\phi(i)}$ of some player $P_{\phi(i)}$~\cite{KER10}. Rank computation is done once in the protocol (step 3). It is required for calculation of the the median $med$, the best-in-class $bic$, the maximum $max$, the bottom quartile $bq$, and the top quartile $tq$. 

Selection is the act of computing the ciphertext of a secret value with specific, i.e. selected, rank~\cite{KER10}. First, $P_S$ chooses a random value $r_i$ individually for each player $P_i$ and computes the ciphertexts $E(x_{\phi(i)}+r_i)$ and $E(r_i)$~\cite{KER10}. That is, the value of $P_i$'s assigned rank blinded by $r_i$ and a $0$ blinded by $r_i$. By using a 1-out-of-2 OT protocol (see Section~\ref{PreliminariesPrimitivesOT}), a player $P_i$ receives $E(x_{\phi(i)}+r_i)$, i.e. the blinded secret value, only if his assigned rank is the one selected~\cite{KER10}. The other players receive the blinded $0$. As these OT steps are identical, we use a template to increase readability of the protocol. Let
\begin{equation}
\begin{split}
OT_{\circ}(m,r_i,p,i) &=\ P_S \xrightarrow{\mathrm{OT}} P_i:\\E^{m}_i&=
\begin{cases}
E\left(x_{\phi(i)} + r_{i}\right) & \text{if}\;\;\; rank_{\phi\left(i\right)} \circ p\\
E\left(r_{i}\right) & \text{otherwise}
\end{cases}
\end{split}
\end{equation}
be the OT step, indexed with a binary comparison operator $\circ$ that takes a statistical measure descriptor $m$, a blinding parameter $r_i$, an array position in the sorted list $p$, and an index $i$. After the OT step, each player rerandomises the value he received by multiplying it by an encrypted $0$ (see Equation~(\ref {eq:PreliminariesHomomorphicEncryptionRerandomisationAddition})) and sends the product to the service provider~\cite{KER10}. The service provider then multiplies the encrypted values he received, removes the random values $r_i$, and gets the ciphertext of $x_{\phi(i)}$~\cite{KER10}. Selection is required for computing the median, best-in-class, maximum, bottom quartile, and the top quartile. It occurs in steps 4 to 6C (OT), steps 10 to 12C (returning the selected values), and steps 15 to 17C (computing the results).

Decryption by the service provider is required since he is supposed to learn the result first~\cite{KER10}. Thus, he is able to round the result before sending it to the players~\cite{KER10}. To decrypt the result $v$, $P_S$ first blinds the result with a random value $r$ and sends the ciphertext $E(v+r)$ to the players~\cite{KER10}. Each player $P_i$ decrypts the blinded result and sends the plaintext $v+r$ together with the corresponding MAC tag
\begin{equation}\label{eq:ApproachesKER10DecryptionMAC}
\theta_i = MAC\left(v+r||i, K_{MAC}\right)
\end{equation}
back to $P_S$~\cite{KER10}. The service provider gets $v$ by subtracting the random value $r$. To prove that he sent the same encrypted, blinded result to every player, $P_S$ computes the hash
\begin{equation}\label{eq:ApproachesKER10DecryptionHash}
\begin{split}
h(&\theta_1=MAC(v+r||1, K_{MAC}),\ldots,\\&\theta_n=MAC(v+r||n, K_{MAC}))
\end{split}
\end{equation}
of the MAC tags $\theta_i$ he received by using a cryptographic hash function~\cite{KER10}. Together with the result $v$, $P_S$ sends this hash to the players. Each $P_i$ then computes the MAC tags and the hash and compares the hash to the one received from the service provider and obtains a validation bit $v_{s_i}$~\cite{KER10}.
This bit, where $s$ indicates the protocol step, is $1$ in case of successful hash validation and $0$ otherwise. It states whether the service provider has sent the same statistical measure to each $P_i$~\cite{KER10}.
Decryption is required for each of the statistical measures mean, variance, median, best-in-class, maximum, bottom quartile, and top quartile~\cite{KER10}. It occurs in steps 2 and 14 to 17C (sending encrypted results), steps 7, 8, and 19 to 26C (returning decrypted, blinded results), steps 9 and 27 to 30C (sending decrypted results), and steps 18 and 31 to 34C (sending the hashed MAC tags).

Based on these preliminaries, our full enhanced protocol is given below in Table~\ref{tab:DesignAdaptedProtocolTable} together with descriptions of the steps we added for the bottom quartile and top quartile computation. These steps are marked with the letter ``B'' and ``C'' in the step label, respectively.
\begin{table*}[t!]
	\renewcommand{\arraystretch}{1.3}
	\centering
	\caption{Enhanced Benchmarking Protocol with Step Labels and Computations}
	\subfloat{%
		\begin{tabular}[t]{p{0.7cm}|p{6.35cm}}
			\hline \textbf{Step} & \textbf{Computation}\\\hline\hline
			1 & \underline{$P_i \rightarrow P_S$:} $E(x_i)$ \\
			\hline
			2 & \underline{$P_S \rightarrow P_i$:} $E(sum+r_1) = E(\sum_{i=1}^n x_i) \cdot E(r_1)$ \\
			3 & $\begin{aligned}[t]\!E(\vec {c_{\phi(i)}}) = &(\ldots, E(c_{\phi(i)_{\phi^\prime(j)}})\\&= E(r_{2_j} \cdot (x_{\phi(i)} - x_{\phi^\prime(j)}) + r_{3_j}),\ldots)\end{aligned}$ \\
			4 & $\mathrm{OT}_{=}(med,r_4,\lceil \frac{n}{2}\rceil,i)$ \\
			5 & $\mathrm{OT}_{\geq}(bic,r_5,\lfloor \frac{3\cdot n}{4}+1\rfloor,i)$ \\
			6 & $\mathrm{OT}_{=}(max,r_6,n,i)$ \\
			6B & $\mathrm{OT}_{=}(bq,r_{6B},\lceil \frac{n}{4}\rceil,i)$ \\
			6C & $\mathrm{OT}_{=}(tq,r_{6C},\lfloor \frac{3\cdot n}{4}+1\rfloor,i)$ \\
			7 & \underline{$P_i \rightarrow P_S$:} $sum+r_1 = D(E(sum+r_1))$ \\
			8 & $MAC(sum+r_1||i,K_{MAC})$ \\
			9 & \underline{$P_S \rightarrow P_i$:} $sum = sum+r_1-r_1$ \\
			10 & \underline{$P_i \rightarrow P_S$:} $E^{med\;\prime}_i = E^{med}_i \cdot E(0)$ \\
			11 & $E^{bic\;\prime}_i = E^{bic}_i \cdot E(0)$ \\
			12 & $E^{max\;\prime}_i = E^{max}_i \cdot E(0)$ \\
			12B & $E^{bq\;\prime}_i = E^{bq}_i \cdot E(0)$ \\
			12C & $E^{tq\;\prime}_i = E^{tq}_i \cdot E(0)$ \\
			13 & $E((x_i-mean)^2) = E((x_i-\frac{sum}{n})^2)$ \\
			\hline
			14 & \underline{$P_S \rightarrow P_i$:} $\begin{aligned}[t]E&(var+r_7)\\&=E(\textstyle\sum_{i=1}^n(x_i-mean)^2) \cdot E(r_7)\\&= (\textstyle\prod_{i=1}^n E((x_i-mean)^2)) \cdot E(r_7)\end{aligned}$ \\
			15 & $E(med+r_8) = (\prod_{i=1}^n E_i^{med\;\prime} \cdot E(-r_{4_i})) \cdot E(r_8)$ \\
			16 & $E(bic+r_9) = (\prod_{i=1}^n E_i^{bic\;\prime} \cdot E(-r_{5_i})) \cdot E(r_9)$ \\
			17 & $E(max+r_{10}) = (\prod_{i=1}^n E_i^{max\;\prime} \cdot E(-r_{6_i})) \cdot E(r_{10})$ \\
			17B & $E(bq+r_{10B}) = (\prod_{i=1}^n E_i^{bq\;\prime} \cdot E(-r_{6B_i})) \cdot E(r_{10B})$ \\
			17C & $E(tq+r_{10C}) = (\prod_{i=1}^n E_i^{tq\;\prime} \cdot E(-r_{6C_i})) \cdot E(r_{10C})$ \\
			18 &$\begin{aligned}[t]h(&MAC(sum+r_1||1,K_{MAC}),\ldots,\\&MAC(sum+r_1||n,K_{MAC}))\end{aligned}$ \\
			19 & \underline{$P_i \rightarrow P_S$:} $var+r_7 = D(E(var+r_7))$ \\
	\end{tabular}}
	\hfil
	\subfloat{%
		\begin{tabular}[t]{p{0.7cm}|p{6.35cm}}
			\hline \textbf{Step} & \textbf{Computation}\\\hline\hline
			20 & $MAC(var+r_7||i,K_{MAC})$	 \\
			21 & $med+r_8 = D(E(med+r_8))$ \\
			22 & $MAC(med+r_8||i,K_{MAC})$ \\
			23 & $bic+r_9 = D(E(bic+r_9))$ \\
			24 & $MAC(bic+r_9||i,K_{MAC})$ \\
			25 & $max+r_{10} = D(E(max+r_{10}))$ \\
			25B & $bq+r_{10B} = D(E(bq+r_{10B}))$ \\
			25C & $tq+r_{10C} = D(E(tq+r_{10C}))$ \\
			26 & $MAC(max+r_{10}||i,K_{MAC})$ \\
			26B & $MAC(bq+r_{10B}||i,K_{MAC})$ \\
			26C & $MAC(tq+r_{10C}||i,K_{MAC})$ \\
			27 & \underline{$P_S \rightarrow P_i$:} $var = var+r_7-r_7$ \\
			28 & $med = med+r_8-r_8$ \\
			29 & $bic = bic+r_9-r_9$ \\
			30 & $max = max+r_{10}-r_{10}$ \\
			30B & $bq = bq+r_{10B}-r_{10B}$ \\
			30C & $tq = tq+r_{10C}-r_{10C}$ \\
			\hline
			31 & \underline{$P_S \rightarrow P_i$:} $\begin{aligned}[t]h(&MAC(var+r_7||1,K_{MAC}),\ldots,\\&MAC(var+r_7||n,K_{MAC}))\end{aligned}$ \\
			32 & $\begin{aligned}[t]h(&MAC(med+r_8||1,K_{MAC}),\ldots,\\&MAC(med+r_8||n,K_{MAC}))\end{aligned}$ \\
			33 & $\begin{aligned}[t]h(&MAC(bic+r_9||1,K_{MAC}),\ldots,\\&MAC(bic+r_9||n,K_{MAC}))\end{aligned}$ \\
			34 & $\begin{aligned}[t]h(&MAC(max+r_{10}||1,K_{MAC}),\ldots,\\&MAC(max+r_{10}||n,K_{MAC}))\end{aligned}$ \\
			34B & $\begin{aligned}[t]h(&MAC(bq+r_{10B}||1,K_{MAC}),\ldots,\\&MAC(bq+r_{10B}||n,K_{MAC}))\end{aligned}$ \\
			34C & $\begin{aligned}[t]h(&MAC(tq+r_{10C}||1,K_{MAC}),\ldots,\\&MAC(tq+r_{10C}||n,K_{MAC}))\end{aligned}$ \\
			\hline
	\end{tabular}}
	\label{tab:DesignAdaptedProtocolTable}
\end{table*}

\paragraph{Round 1 (step 1)}\label{ConceptProtocolRound1}
Each player $P_i$ sends his encrypted input to the service provider $P_S$.

\paragraph{Round 2 (steps 2-13)}\label{ConceptProtocolRound2}
The service provider computes the encrypted, blinded sum of the input values and sends it to the players $P_i$. Furthermore, $P_S$ conducts a rank computation after which each player has the rank of some player $P_j$'s input value. Given that rank, each $P_i$ receives either an encrypted, blinded statistical measure or an encrypted random value via OT depending on whether his assigned rank fits the respective selection criterion. This is repeated for each of the statistical measures median, best-in-class, maximum, bottom quartile, and top quartile. Afterwards, the players decrypt the blinded sum they received and send it back to $P_S$ together with a MAC tag of the blinded sum. Furthermore, $P_S$ sends the sum to each $P_i$. Then, each player rerandomises his OT step outputs and sends them back to the service provider. Then each player computes the squared difference between his input and the mean and sends the encrypted result to $P_S$ as the basis for the variance computation.

Steps 6B and 6C are OT steps that are part of the selection of the bottom quartile and top quartile values of the sorted list of inputs. In step 6B, the selection criterion of the OT protocol $rank_{\phi\left(i\right)} = \lceil \frac{n}{4}\rceil$ is the index of the sorted list's bottom quartile element. For step 6C, the top quartile index $\lfloor \frac{3\cdot n}{4} +1 \rfloor$ is used. 
Steps 12B and 12C rerandomise the ciphertext received from the service provider in steps 6B and 6C.

\paragraph{Round 3 (steps 14-30C)}\label{ConceptProtocolRound3}
The service provider computes the encrypted, blinded statistical measures variance, median, best-in-class, maximum, bottom quartile, and top quartile by multiplying the values received in round 2. He sends them to the players together with the hashed MAC tags of the blinded sum. The latter is then used by the players to validate whether each player previously received the same blinded sum. Similar to round 2, each player then decrypts the blinded statistical measures and sends them to $P_S$ together with the respective MAC tags. In the last steps of round 3, $P_S$ sends the unblinded statistical measures to the players.

Steps 17B and 17C return the encrypted, rank-based statistical measures to the players. They are computed by multiplying the $n$ encrypted, rerandomised OT messages that $P_S$ received in steps 12B and 12C. Additionally, the statistical measures are blinded by multiplying the product by an encrypted random value $r_{10B}$ and $r_{10C}$, respectively.
In steps 25B and 25C, each player decrypts the blinded statistical measures he received in steps 17B and 17C and sends the result to $P_S$.
In steps 26B and 26C, each player sends the MAC tags of the blinded bottom quartile and top quartile, respectively. The blinded statistical measures are concatenated with player $P_i$'s index $i$ before computing the corresponding MAC tag. This MAC computation requires the symmetric MAC key $K_{MAC}$.
Steps 30B and 30C are used by $P_S$ to send the output, i.e. the decrypted, unblinded statistical measures, to the players.

\paragraph{Round 4 (steps 31-34C)}\label{ConceptProtocolRound4}
The service provider sends to each player the hashed MAC tags of the blinded statistical measures variance, median, best-in-class, maximum, bottom quartile, and top quartile. These are used by the players for validation of output integrity.

Steps 34B and 34C are used by the service provider to distribute the hashed MAC tags of the statistical measures. Step 34B is the hash of the $n$ bottom quartile MAC tags that the players sent to $P_S$ in step 26B. Step 34C in turn is the hash of the $n$ top quartile MAC tags of step 26C.

\subsection{Implementation}\label{DesignImplementation}

\begin{figure*}[t!]
	\centering
	\includegraphics[scale=0.7]{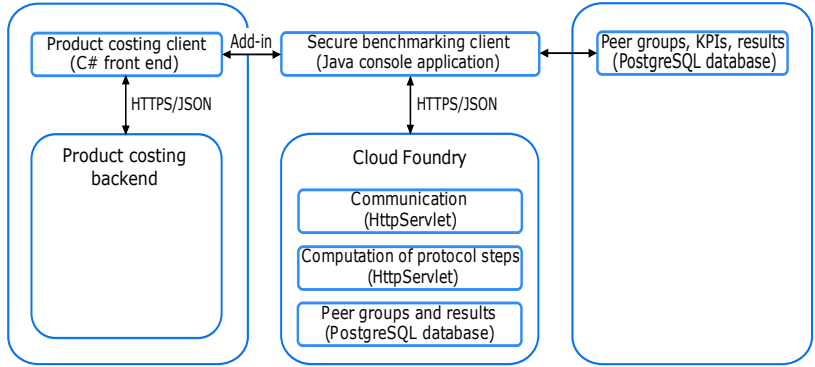}
	\caption{Block Diagram of the Prototype's Architecture}
	\label{fig:DesignPrototypeBlockDiagram}
\end{figure*}

Our prototype consists of two main parts: the secure benchmarking client and the secure benchmarking service. Both are written in Java and have their own PostgreSQL database to enable persistent data storage. The service is a Java servlet, running on Cloud Foundry (see~\cite{CFF17}), while the client is a Java console application. Additionally, we developed a C\# front end add-in for the product costing suite. It utilises the client implementation to execute the benchmarking protocol for actual product costing key figures. Client and service communicate via HTTPS sending JSON strings.
An overview of this secure benchmarking system is depicted in Fig.~\ref{fig:DesignPrototypeBlockDiagram}.
\section{Evaluation}\label{Evaluation}

This Section provides an evaluation of the adapted approach implemented in the prototype. The evaluation refers to the requirements that are described in Section~\ref{Introduction}.

\subsection{Statistical measures}\label{EvaluationStatisticalMeasures}

The selected protocol of~\cite{KER10} computes the statistical measures mean, variance, median, maximum, and best-in-class while the adapted protocol additionally computes the bottom quartile and top quartile.
The correctness of the adapted protocol has been proven (see Appendix~\ref{ConceptProtocolAdaptationsCorrectness}).
Therefore, the adapted protocol fulfils the requirements for necessary and optional statistical measures in the context of product costing.

\subsection{Communication}\label{EvaluationCommunication}

As described in Section~\ref{ApproachesKER10}, the protocol of~\cite{KER10} is based on the service provider model. The same applies to the adapted protocol since our adaptations presented in Section~\ref{DesignAdaptedProtocol} only add further computation steps but do not affect the communication. The service provider's implementation of the protocol runs as a Cloud Foundry service (see Section~\ref{DesignImplementation}) proving the protocol's cloud suitability. No pairwise communication between players takes place. Messages sent from $P_i$ to $P_S$ and vice versa do not contain any private data of some player $P_{j\neq i}$ that allows for an identification of $P_j$. This enables anonymity among the players in case of a non-colluding \(P_S\).

\subsection{Performance}\label{EvaluationPerformance}

As stated in Section~\ref{ApproachesSummary}, the quadratic complexity of the selected approach may prove critical for large $n$. Therefore, we conducted an extensive evaluation of the prototype's performance. We examined the computational and communication complexity of our protocol both theoretically and empirically.

\subsubsection{Theoretical performance of the adapted protocol}\label{EvaluationPerformanceTheoretical}

The most expensive part of the original benchmarking protocol is the rank computation in step 3. 
Its computational and communication complexity is quadratic in the number of players $n$, i.e. $\mathcal{O}(n^2)$~\cite{KER10}. This also holds for the rank computation of the adapted protocol since step 3 is not affected by the adaptations.
Since the newly added steps each have complexity that is below $\mathcal{O}(n^2)$, the total computational and communication complexity of the protocol is $\mathcal{O}(n^2)$. The computational and communication complexity of each step of the adapted protocol is given in Appendix~\ref{AppendixComplexity}.

\subsubsection{Practical performance of the prototype}\label{EvaluationPerformancePractical}

To compare the prototype's performance to the theoretical complexity of the adapted protocol, a practical evaluation is performed in this Section. For this evaluation, a number of benchmarks were executed while the net execution time was measured. The net execution time is the time required for one entire protocol execution minus the time during which an inactive participant delays the execution, causing the others to wait. 
These benchmarks were conducted in different scenarios to determine the influence of the parameters peer group size $n$, network latency, bandwidth, and asymmetric (Paillier) key length. Furthermore, two worst case scenarios were considered combining several of these parameters.

For the empirical analysis, the Java servlet 
was uploaded to a Cloud Foundry trial instance. 
The $n$ clients were run in the form of the Java console application on three local client notebooks. 
Preliminary analyses indicated hardware limitations on the part of the service provider and the client machines. Hence, the performance evaluation was conducted for peer groups of $n\leq 60$ players and extrapolated for larger $n$. 
These $n$ clients were equally distributed among the three PCs. In each of the corresponding tests, the net execution time $t$ was measured for a default Paillier key length of $768$ bits, a default bandwidth of circa $17$ Mbit/s, and a network latency to the service provider of about $6$ milliseconds.

\paragraph{Peer group size}\label{EvaluationPerformancePracticalEvaluationPeergroupSize}

\begin{figure*}[t!]
	\centering
	\centerline{
		\subfloat[Different Peer Group Sizes, Extrapolated]{\includegraphics[width=2.24in]{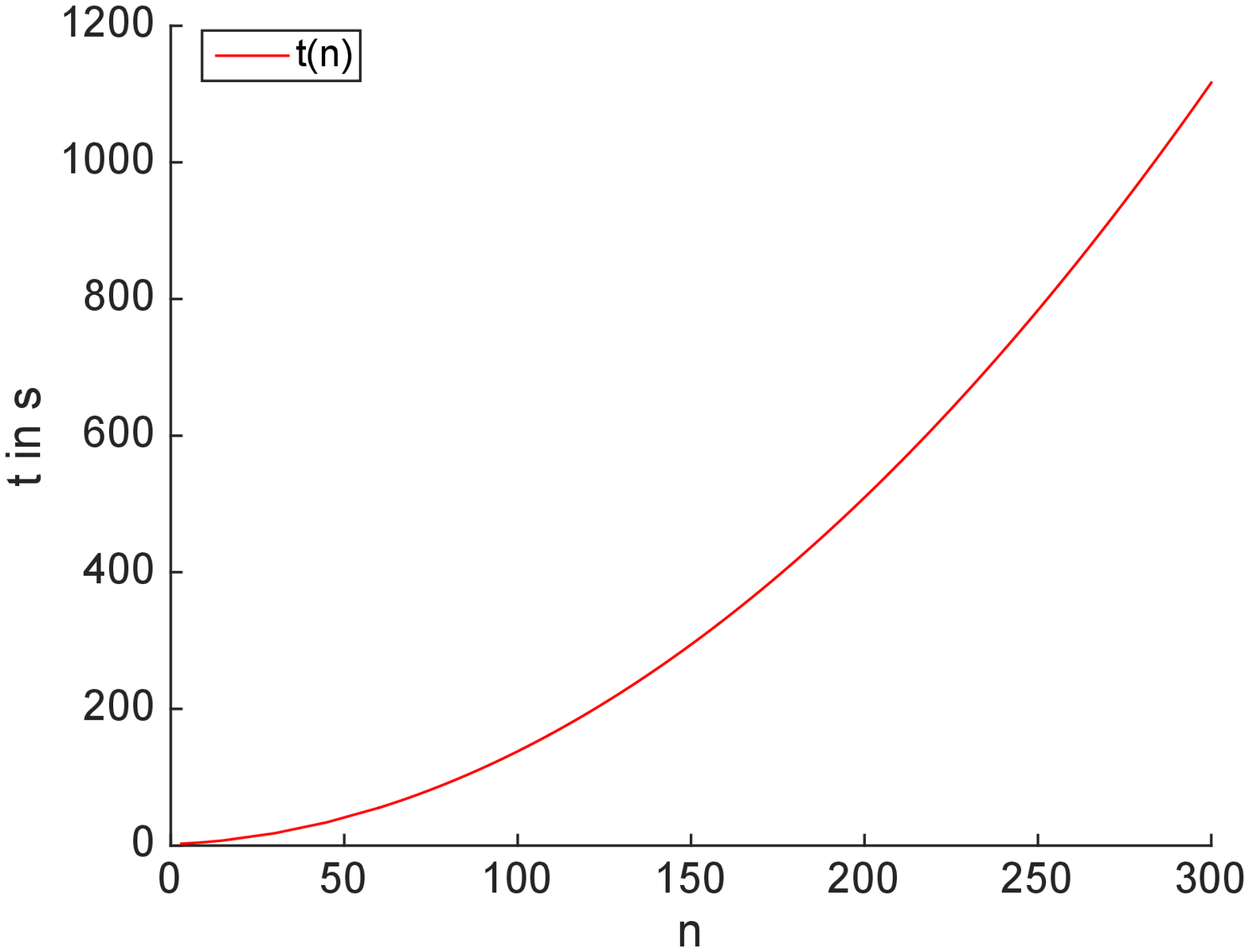}%
			\label{fig:EvaluationPeerGroupSize}}
		\hfil
		\subfloat[Worst Case 1, Extrapolated]{\includegraphics[width=2.24in]{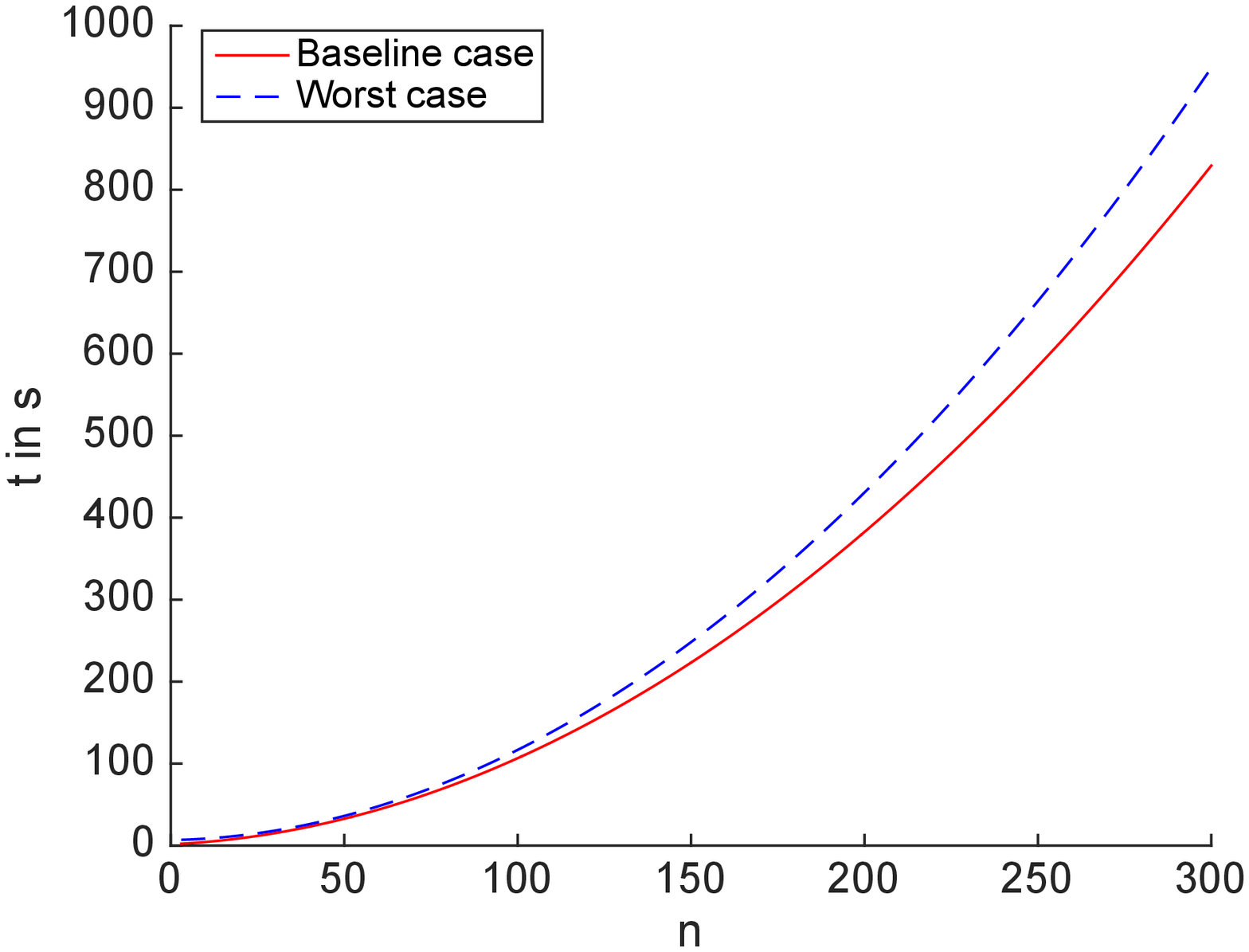}%
			\label{fig:EvaluationWorstCase1Estimated}}
		\hfil
		\subfloat[Worst Case 2, Extrapolated]{\includegraphics[width=2.24in]{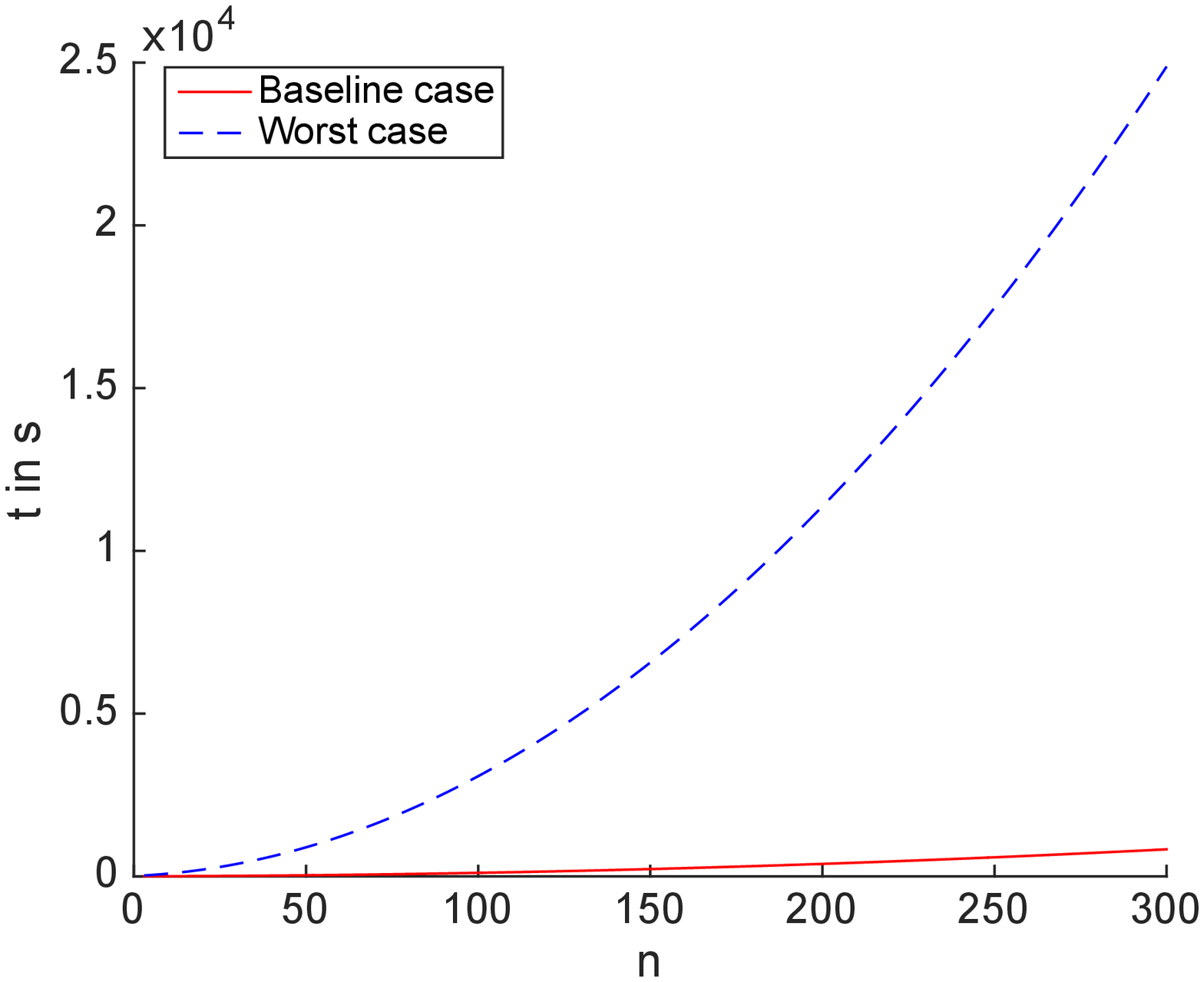}%
			\label{fig:EvaluationWorstCase2Estimated}}}
	\caption{Net Execution Times of the Enhanced Benchmarking Protocol}
\end{figure*}

Fig.~\ref{fig:EvaluationPeerGroupSize} depicts the extrapolated net execution time for peer group sizes of $n\leq300$ players given the measured net execution time for $n\leq60$ players.
With circa $1.200$ seconds, the net execution time was below the required maximum of $24$ hours. For the remaining parameters network latency, bandwidth, and key length, the default values were used. 
This scenario serves as the baseline scenario for the remaining analyses of the performance evaluation.

\paragraph{Network latency}\label{EvaluationPerformancePracticalEvaluationNetworkLatency}

To examine the effect of the network latency, the net execution time was measured for the default network latency of $6$ milliseconds as well as for this default value plus an offset. The network latency offset had to be suitable for simulating global communication. 
Based on the network latencies for different continents as well as for intercontinental communication given in~\cite{VER17}, an average offset of $150$ milliseconds was chosen. 
Compared to the baseline, the offset only significantly affects the net execution time for smaller peer groups of $n\leq 40$ players. With increasing peer group size $n$, the effect of the network latency decreases.

\paragraph{Bandwidth}\label{EvaluationPerformancePracticalEvaluationBandwidth}

Customers can reasonably be assumed to have access to a broadband network, which is defined as a network with a transmission capacity of at least $2$ Mbit/s~\cite{ITU11}.
Even for such a low bandwidth, the difference between the net execution times was small. Increasing the bandwidth by a factor of $8.5$ barely affected the net execution time.

\paragraph{Key length}\label{EvaluationPerformancePracticalEvaluationKeyLength}

For investigating the effect of the key length on the net execution time, the key lengths $768$, $1024$, $1536$, and $2048$ bits were taken into consideration.
These investigations showed two facts. On the one hand, the net execution time increased disproportionately with growing key length. This is due to the complexity of the modular exponentiation of Paillier's cryptosystem, which is cubic in the key length~\cite{KER10}. On the other hand, for keys of $2048$ bits, the net execution time for $n=300$ was about $25.000$ seconds, i.e. approximately $7$ hours. 

\paragraph{Worst case}\label{EvaluationPerformancePracticalEvaluationWorstCase}

For the worst case analysis, two scenarios are taken into consideration. This analysis refers to the worst case from a net execution time perspective, not from a security perspective. The first scenario investigates the combined effect of the peer group size, network latency, and bandwidth while the second worst case scenario also takes the key length into consideration.
The extrapolated net execution time for the first worst case scenario is depicted in Fig.~\ref{fig:EvaluationWorstCase1Estimated}. 
For growing peer group size, the fitted curves diverge. 
The net execution time for $n=300$ was below the required maximum of $24$ hours. 
The fitted curves for both cases grow faster than they diverge. Hence, the influence of the computations on the net execution time can be assumed to be higher than the impact of the communication.
The extrapolated net execution time of the second worst case scenario is depicted in Fig.~\ref{fig:EvaluationWorstCase2Estimated}. Due to the large difference in the key length, which was $768$ bits in the baseline case and $2048$ bits in the worst case, the fitted curves grow very differently. However, both grow quadratically. 
The net execution time of the second scenario was circa $7$ hours. Hence, the prototype's performance meets the requirements even for larger keys of $2048$ bits, poor internet connection, and peer groups of $n\leq300$ players. However, with its quadratic complexity, the net execution time of this system would soon exceed the desired maximum of $24$ hours for larger peer groups. This is mostly due to the extensive rank computation.

\subsection{Security}\label{EvaluationSecurity}

We proved the adapted protocol to be secure against computationally bounded semi-honest adversaries (see Appendix~\ref{ConceptProtocolAdaptationsPrivacy}). This privacy proof shows that up to $n-1$ players can be corrupted without them learning anything about the non-corrupted players' inputs that cannot be inferred from the corrupted players' inputs and the output.
If the service provider himself is corrupted, privacy of the inputs ensures that he learns nothing about the players’ secret values as long as none of the players is corrupted at the same time~\cite{KER10}. The possibility of a player eavesdropping on the communication between another player and the service provider is precluded as they communicate over pairwise secure channels.
Integrity of the output is enabled via cryptographic hashes and MACs. 
Further security related properties, such as secure data storage, access control, availability of the service, and robustness regarding players dropping out of a protocol execution, are beyond the scope of this paper.

\subsection{Summary of the evaluation}\label{EvaluationSummary}

The statistical measures provided by the implemented protocol are the mean, variance, median, maximum, best-in-class, bottom quartile, and the top quartile, and the best-in-class.
The prototype implements the service provider model, enabling anonymity among the players as well as cloud suitability. 
Our enhanced secure benchmarking protocol has been proven to ensure confidentiality of the players' secret inputs against computationally bounded semi-honest adversaries.
Output integrity is also provided.
The computational and communication complexity both are $\mathcal{O}(n^2)$. The net execution time for $300$ players is circa $7$ hours.
Consequently, the prototype implementing the adapted protocol meets each of the requirements defined in Section~\ref{Introduction}. It even provides additional statistical measures. However, our extensive evaluation showed the bottleneck of our protocol, which is the quadratic rank computation.
\section{Discussion}\label{Discussion}

A variety of attacks that are possible for TTP-based benchmarking analyses can also be applied to our system. This includes inference attacks, where an adversary-controlled player $P_i$ repeatedly runs benchmarks for the same KPI and peer group, each time with a different input $x_i$, in order to gain additional knowledge of other players' secret inputs. Such attacks can be precluded via organisational measures like retention periods. In the described product cost optimisation scenario, benchmarks are not supposed to be executed on demand but rather once every one or two quarters on pre-defined dates. Given the long period of time between two benchmarks and assuming that at least some players' KPIs changed during that period, the value of such an attack would be rather limited.

Attacks where a player inputs an incorrect KPI to temper with the results are much harder to circumvent as they require a mechanism for analysing the semantics of the inputs. However, as this is immanent in benchmarking in general, such behaviour was beyond the scope of our considerations.
\section{Conclusion and Future Work}\label{Conclusion}

This paper provides an overview of benchmarking based on secure multi-party computation. It elaborates requirements for a secure benchmarking system in the context of product cost optimisation. Based on these requirements, generic approaches to secure KPI benchmarking were discussed, the existing related work was reviewed, and fitting approaches were described. The most suitable approach was selected, extended, implemented in a prototype, and extensively evaluated. The resulting protocol meets all of the requirements for a privacy-preserving benchmarking system in the given context of product cost optimisation. However, its computational and communication complexity that is quadratic in the number of players $n$, i.e. $\mathcal{O}(n^2)$, shows potential for further optimisation.

In our future work, we will focus on reducing the computational complexity by simplifying the rank computation. In the presented protocol, this sorting step compares each player's value to any other player's value, causing a quadratic number of comparisons. Fortunately, the lower bound for comparisons in sorting is $\mathcal{O}(n \cdot \log n)$~\cite{CLRS09}. Such a sorting mechanism with less comparisons would likely need to be more interactive, causing a higher communication complexity.
Furthermore, it would require an additional step for hiding the order of the inputs. Otherwise, the service provider would learn the players' ranks in the sorted list of inputs.

\printbibliography
\newpage\clearpage
\appendix

\section{Proof sketches}\label{AppendixProofSketches}

\subsection{Correctness of the adapted protocol}\label{ConceptProtocolAdaptationsCorrectness}


\subsubsection{Bottom quartile}\label{ConceptProtocolAdaptationsCorrectnessBottomQuartile}

Assume the original protocol as presented in~\cite{KER10} is correct. Consequently, the rank set computed in step 3 is a set, not a multiset, containing the $n$ consecutive integers $1,...,n$~\cite{KER10}.
The bottom quartile is the element at position (rank) $\lceil \frac{n}{4}\rceil$ of the sorted list of inputs.
Step 6B yields
\begin{equation}\label{eq:ConceptProofCorrectness6B2}
E^{bq}_i = E\left(x_{\phi(i)} + r_{6B_i}\right)
\end{equation}
for the one player $P_i$ with rank $rank_{\phi\left(i\right)} = \lceil \frac{n}{4}\rceil$ and
\begin{equation}\label{eq:ConceptProofCorrectness6B1}
E^{bq}_i = E\left(r_{6B_i}\right)
\end{equation}
for each of the remaining $n-1$ players $P_i$ with rank $rank_{\phi\left(i\right)} \neq \lceil \frac{n}{4}\rceil$.
In step 12B, each player's $E^{bq}_i$ is rerandomised by multiplying it with the encrypted identity element $0$. This does not affect the corresponding plaintext, which is
\begin{equation}\label{eq:ConceptProofCorrectness12B1}
\begin{aligned}
D\left(E^{bq\;\prime}_i\right) &= D\left(E^{bq}_i \cdot E\left(0\right)\right)\\&= D\left(E\left(x_{\phi(i)} + r_{6B_i} + 0\right)\right)\\&= D\left(E\left(x_{\phi(i)} + r_{6B_i}\right)\right)\\&= D\left(E^{bq}_i\right)\\&= x_{\phi(i)} + r_{6B_i}
\end{aligned}
\end{equation}
for the one player $P_i$ with rank $rank_{\phi\left(i\right)} = \lceil \frac{n}{4}\rceil$ and
\begin{equation}\label{eq:ConceptProofCorrectness12B2}
\begin{aligned}
D\left(E^{bq\;\prime}_i\right) &= D\left(E^{bq}_i \cdot E\left(0\right)\right)\\&= D\left(E\left(r_{6B_i} + 0\right)\right)\\&= D\left(E\left(r_{6B_i}\right)\right)\\&= D\left(E^{bq}_i\right)\\&= r_{6B_i}
\end{aligned}
\end{equation}
for each of the remaining $n-1$ players $P_i$ with rank $rank_{\phi\left(i\right)} \neq \lceil \frac{n}{4}\rceil$ (see Equation~(\ref{eq:PreliminariesHomomorphicEncryptionRerandomisationAddition})).
In step 17B, the service provider $P_S$ computes the product of the $n$ rerandomised ciphertexts $E^{bq\;\prime}_i$ each multiplied by the corresponding encrypted, negated random value $r_{6B_i}$. According to Equation~(\ref{eq:PreliminariesHomomorphicEncryptionAdditionMult}), this yields
\begin{equation}\label{eq:ConceptProofCorrectness17B1}
\begin{aligned}
\prod_{i=1}^n &E_i^{bq\;\prime} \cdot E\left(-r_{6B_i}\right)\\&= E\left(r_{6B_1} + \cdots + r_{6B_n} + x_{\phi(i)} - r_{6B_1} - \cdots - r_{6B_n}\right)\\&= E\left(x_{\phi(i)}\right) = E\left(bq\right).
\end{aligned}
\end{equation}
Furthermore, this product of ciphertexts is multiplied by an encrypted random value $r_{10B}$ resulting in the encrypted, blinded statistical measure
\begin{equation}\label{eq:ConceptProofCorrectness17B2}
\begin{aligned}
E\left(bq\right) \cdot E\left(r_{10B}\right) = E\left(bq + r_{10B}\right).
\end{aligned}
\end{equation}
Decrypting this ciphertext in step 25B yields the decrypted, blinded statistical measure
\begin{equation}\label{eq:ConceptProofCorrectness25B}
\begin{aligned}
D\left(E\left(bq + r_{10B}\right)\right) = bq + r_{10B}.
\end{aligned}
\end{equation}
The subtraction of $r_{10B}$ by $P_S$ in step 30B results in the statistical measure
\begin{equation}\label{eq:ConceptProofCorrectness30B}
\begin{aligned}
bq + r_{10B} - r_{10B} = bq.
\end{aligned}
\end{equation}
Assume that all the players $P_i$ and the service provider $P_S$ use the same MAC function $MAC(\cdot)$ and the same cryptographic hash function $h(\cdot)$. Furthermore, assume that the symmetric key of the MAC function is known to every player $P_i$ but not to $P_S$. Equality of the two hashes computed in 34B given the MAC tags of steps 26B proves integrity of the statistical measure bottom quartile~\cite{KER10}.
\\
This completes the proof of correctness of the steps related to the bottom quartile computation.~$\hfill\square$
\\

\subsubsection{Top quartile}\label{ConceptProtocolAdaptationsCorrectnessTopQuartile}

The correctness proof of the top quartile computation is very similar to the one of the bottom quartile computation. The only difference apart from step labels and variable names is the selection criterion of the OT step 6C, i.e. $\lfloor \frac{3\cdot n}{4} +1 \rfloor$, which is the position of the top quartile element.

\subsection{Privacy of the adapted protocol}\label{ConceptProtocolAdaptationsPrivacy}

For our protocol to be secure in the semi-honest model, it has to be provided that anything an adversary can learn during a protocol execution can as well be learned only from the input and the output of the protocol~\cite{KER10}. To prove that, one needs to show that the view $V$ of an adversary can be simulated only based on the input and output~\cite{KER10}.
The protocol privately computes 
the respective quartile if a simulator $S$ is able to generate a view $V^\prime$ that is computationally indistinguishable from a participant’s view $V$~\cite{KER10}. This simulator creates the protocol input and the coin tosses himself. The former can be done by taking the original input while the latter is simulated by using the same pseudorandom generator (PRG) that is used for generating the random numbers in the protocol. Therefore, only the messages $m_i$ that the participant receives are relevant~\cite{KER10}. For these, the simulator has to generate a message $m_i^\prime$ for each message $m_i$ of the view $V$ such that both are computationally indistinguishable.

\subsubsection{The players' view}\label{ConceptProtocolAdaptationsPrivacyPlayersView}

Similar to the proof of the original protocol in~\cite{KER10}, the OT protocol is substituted by its ideal functionality for simplification purposes. Since the players have the secret decryption key, in case the decrypted messages can be simulated by $S_{P_i}$ computationally indistinguishable, the same applies to the encrypted messages. This is true since encryption can be regarded as a deterministic mapping of probability distributions~\cite{KER10}.
During a protocol execution, player $P_i$ receives the following messages. The arrow ``$\rightarrow$'' shows the corresponding plaintext that $P_i$ can compute himself.
\begin{itemize}
	\addtolength{\itemindent}{0.25cm}
	\item[6B.] $E^{bq}_i =
	\begin{cases}
	E\left(x_{\phi(i)} + r_{6B_i}\right) \rightarrow x_{\phi(i)} + r_{6B_i}\\
	E\left(r_{6B_i}\right) \rightarrow r_{6B_i}
	\end{cases}$
	\item[6C.] $E^{tq}_i =
	\begin{cases}
	E\left(x_{\phi(i)} + r_{6C_i}\right) \rightarrow x_{\phi(i)} + r_{6C_i}\\
	E\left(r_{6C_i}\right) \rightarrow r_{6C_i}
	\end{cases}$
	\item[17B.] $E\left(bq+r_{10B}\right) \rightarrow bq+r_{10B}$
	\item[17C.] $E\left(tq+r_{10C}\right) \rightarrow tq+r_{10C}$
	\item[30B.] $bq$
	\item[30C.] $tq$
	\item[34B.] $\begin{aligned}[t]h(&MAC(bq+r_{10B}||1,K_{MAC}),...,\\&MAC(bq+r_{10B}||n,K_{MAC}))\end{aligned}$
	\item[34C.] $\begin{aligned}[t]h(&MAC(tq+r_{10C}||1,K_{MAC}),...,\\&MAC(tq+r_{10C}||n,K_{MAC}))\end{aligned}$
\end{itemize}

\subsubsection{The service provider's view}\label{ConceptProtocolAdaptationsPrivacyServiceProvidersView}

Different from the players, the service provider $P_S$ does not have an input. His output are the statistical measures $bq$ and $tq$. Since he does not have access to the secret decryption key, he cannot decrypt encrypted messages~\cite{KER10}. The messages he receives are as follows.
\begin{itemize}
	\addtolength{\itemindent}{0.25cm}
	\item[12B.] $E^{bq\;\prime}_i = E^{bq}_i \cdot E\left(0\right)$
	\item[12C.] $E^{tq\;\prime}_i = E^{tq}_i \cdot E\left(0\right)$
	\item[25B.] $bq+r_{10B}$
	\item[25C.] $tq+r_{10C}$
	\item[26B.] $MAC\left(bq+r_{10B}||i,K_{MAC}\right)$
	\item[26C.] $MAC\left(tq+r_{10C}||i,K_{MAC}\right)$
\end{itemize}

\subsubsection{The service provider's and the players' simulators}\label{ConceptProtocolAdaptationsPrivacySimulators}

Given that $dom(\cdot)$ denotes the domain of a function, the simulator $S_{P_i}$ for a player $P_i$ generates the following simulated messages.
The names of the random values are not related to those of the random values used in the protocol description.
\begin{itemize}
	\addtolength{\itemindent}{0.25cm}
	\item[6B.] A random value $r_1$, uniformly chosen from $dom(D(\cdot))$
	\item[6C.] A random value $r_2$, uniformly chosen from $dom(D(\cdot))$
	\item[17B.] A random value $r_3$, uniformly chosen from $dom(D(\cdot))$
	\item[17C.] A random value $r_4$, uniformly chosen from $dom(D(\cdot))$
	\item[30B.] $bq$
	\item[30C.] $tq$
	\item[34B.] If the validation bit $v_{34B_i}$ equals $1$, the simulator's value is\\
	$\begin{aligned}h(&MAC(bq+r_3||1,K_{MAC}),...,\\&MAC(bq+r_3||n,K_{MAC}))\end{aligned}$\\otherwise it is a random value $r_5$, uniformly chosen from $dom(h(\cdot))$
	\item[34C.] If the validation bit $v_{34C_i}$ equals $1$, the simulator's value is\\ $\begin{aligned}h(&MAC(tq+r_4||1,K_{MAC}),...,\\&MAC(tq+r_4||n,K_{MAC}))\end{aligned}$\\
	otherwise it is a random value $r_6$, uniformly chosen from $dom(h(\cdot))$
\end{itemize}

The corresponding simulator $S_{P_S}$ for the service provider $P_S$ generates the following simulated messages.
\begin{itemize}
	\addtolength{\itemindent}{0.25cm}
	\item[12B.] A random value $r_7$, uniformly chosen from $dom(E(\cdot))$
	\item[12C.] A random value $r_8$, uniformly chosen from $dom(E(\cdot))$
	\item[25B.] $bq+r_9$ where $r_9$ is an internal coin toss
	\item[25C.] $tq+r_{10}$ where $r_{10}$ is an internal coin toss
	\item[26B.] A random value $r_{11}$, uniformly chosen from $dom(MAC(\cdot))$
	\item[26C.] A random value $r_{12}$, uniformly chosen from $dom(MAC(\cdot))$
\end{itemize}

\setcounter{footnote}{0}
\renewcommand*{\thefootnote}{\alph{footnote}}

\begin{table*}[!t]
	\renewcommand{\arraystretch}{1.3}
	\caption{Computational and Communication Complexity of the Enhanced Protocol per Participant}
	
	\centering
	\begin{tabular}{l|c|c|c|c|c|c|c}
		\hline
		
		\textbf{Step} & \textbf{$E(\cdot)$} & \textbf{$D(\cdot)$} & \textbf{Exp} & \textbf{Mult} & \textbf{Add} & \textbf{Inv} & \textbf{Values to send}\\\hline\hline
		
		\textbf{1} & $1$ & & & & & & $1$\\\hline
		
		\textbf{2} & $1$ & & & $n$ & & & $n$\\\hline
		
		\textbf{3} & $n^2$ & & $n^2$ & $2\!\cdot\!n^2$ & & $n^2$ & $n^2$\\\hline
		
		\textbf{4-6C ($P_i$)} & $1$ & & & & $2$ & & $1$\\\hline
		
		\textbf{4-6C ($P_S$)} & $n$ & & & $n$ & & & $n$\\\hline
		
		\textbf{7} & & $1$ & & & & & $1$\\\hline
		
		\textbf{8} & & & \footnoteref{ftn:AppendixComplexityDependsMAC} & \footnoteref{ftn:AppendixComplexityDependsMAC} & \footnoteref{ftn:AppendixComplexityDependsMAC} & & $1$\\\hline
		
		\textbf{9} & & & & & $1$ & & $n$\\\hline
		
		\textbf{10-12C} & $1$ & & & $1$ & & & $1$\\\hline
		
		\textbf{13} & $1$ & & & $2$ & $1$ & & $1$\\\hline
		
		\textbf{14} & $1$ & & & $n$ & & & $n$\\\hline
		
		\textbf{15-17C} & $n+1$ & & & $2\!\cdot\!n$ & & & $n$\\\hline
		
		\textbf{18} & & & \footnoteref{ftn:AppendixComplexityDependsHash} & \footnoteref{ftn:AppendixComplexityDependsHash} & \footnoteref{ftn:AppendixComplexityDependsHash} & & $n$\\\hline
		
		\textbf{19, 21, 23, 25, 25B, 25C} & & $1$ & & & & & $1$\\\hline
		
		\textbf{20, 22, 24, 26, 26B, 26C} & & & \footnoteref{ftn:AppendixComplexityDependsMAC} & \footnoteref{ftn:AppendixComplexityDependsMAC} & \footnoteref{ftn:AppendixComplexityDependsMAC} & & $1$\\\hline
		
		\textbf{27-30C} & & & & & $1$ & & $n$\\\hline
		
		\textbf{31-34C} & & & \footnoteref{ftn:AppendixComplexityDependsHash} & \footnoteref{ftn:AppendixComplexityDependsHash} & \footnoteref{ftn:AppendixComplexityDependsHash} & & $n$\\\hline\hline
		
		\textbf{Total $\left(P_i\right)$} & \textbf{$8$} & \textbf{$7$} & \footnoteref{ftn:AppendixComplexityDependsMAC} & \textbf{$7$}\footnoteref{ftn:AppendixComplexityDependsMAC} & \textbf{$11$}\footnoteref{ftn:AppendixComplexityDependsMAC} & & \textbf{$26$}\\\hline
		
		\textbf{Total $\left(P_S\right)$} & \textbf{$n^2\!+\!10\!\cdot\!n\!+\!7$} & \textbf{$0$} & \textbf{$n^2$}\footnoteref{ftn:AppendixComplexityDependsHash} & \textbf{$2\!\cdot\!n^2\!+\!17\!\cdot\!n$}\footnoteref{ftn:AppendixComplexityDependsHash} & \textbf{$7$}\footnoteref{ftn:AppendixComplexityDependsHash} & \textbf{$n^2$} & \textbf{$n^2\!+\!26\!\cdot\!n$}\\\hline
	\end{tabular}
	\vspace{0.2in}
	\label{tab:AppendixComplexityComputational}
\end{table*}

\subsubsection{Comparison}\label{ConceptProtocolAdaptationsPrivacyComparison}

Privacy is proven if the simulator generates an output that is computationally indistinguishable from a participant's view. To prove computational indistinguishability, it is sufficient to show that two functions are identically distributed~\cite{KER10}.

In case of steps 6B, 6C, 17B, and 17C, the secret values are blinded by a random number that is added to it. Consequently, these sums are identically distributed to uniformly chosen random numbers~\cite{KER10}. Therefore, they are computationally indistinguishable from the uniformly chosen random numbers provided by the simulator $S_{P_i}$~\cite{KER10}.

The messages of steps 12B and 12C are rerandomised encrypted values that cannot be decrypted by the receiver. According to the proof of semantic security of Paillier's cryptosystem, the ciphertexts are computationally indistinguishable from uniformly chosen random numbers~\cite{KER10}. Therefore, they are also computationally indistinguishable from the corresponding output provided by the simulator $S_{P_S}$~\cite{KER10}.

In steps 25B and 25C, the messages are blinded statistical measures, that is, sums of a statistical measure and a uniformly chosen random value. The statistical measure is part of the receiver's, i.e. the service provider's, output. Therefore, the simulator can copy it from the output~\cite{KER10}. The random variable was previously generated by $P_S$ himself and is hence known to him~\cite{KER10}. Given the same PRG, the simulator can generate a random value which is identically distributed~\cite{KER10}. Therefore, the messages and $S_{P_S}$'s output are identically distributed and hence computationally indistinguishable~\cite{KER10}.

Assuming that the service provider is not capable of MAC forgery and does not know the secret MAC key $K_{MAC}$, MAC tags are computationally indistinguishable from random numbers for him~\cite{KER10}. Therefore, the messages of steps 26B and 26C are computationally indistinguishable from the uniformly chosen random numbers provided by the simulator $S_{P_S}$.

The simulation of steps 30B and 30C only requires copying $P_i$'s outputs, i.e. the statistical measures bottom quartile and top quartile, that are known to the simulator. The messages of player $P_i$ and the output of the simulator $S_{P_i}$ are therefore computationally indistinguishable~\cite{KER10}.

In steps 34B and 34C, the hashes sent from the service provider can either match the ones computed by player $P_i$ or not. If they do not match, the received hash appears as a uniformly chosen random number to $P_i$ as he cannot compute a pre-image of the hash~\cite{KER10}. If the hashes match, the simulator's output is the actual hash~\cite{KER10}. In both cases, the simulator's outputs and the player's view are identically distributed and therefore computationally indistinguishable~\cite{KER10}.
\\
This completes the proof of privacy of the adapted protocol.~$\hfill\square$

\subsection{Computational and communication complexity of the adapted protocol}\label{AppendixComplexity}

For each step, the number of required encryptions, decryptions, exponentiations, multiplications, and additions as well as the number of modular inversions is given in Table~\ref{tab:AppendixComplexityComputational}. Together they form the computational complexity. The communication complexity is given in the rightmost column. Consecutive steps with the same complexity, such as steps 4 to 6C, are shown in a single row of the same table.
\pagebreak
The lower-case letters in this table do not represent variables but instead correspond to the footnotes~\footnote{Depends on the MAC function being used\label{ftn:AppendixComplexityDependsMAC}}~and~\footnote{Depends on the hash function being used\label{ftn:AppendixComplexityDependsHash}}.

\end{document}